\documentclass[conference]{IEEEtran}
\IEEEoverridecommandlockouts
\usepackage{cite}
\usepackage{amsmath,amssymb,amsfonts}
\usepackage{algorithmic}
\usepackage{graphicx}
\usepackage{textcomp}
\usepackage{xcolor}
\usepackage{subcaption}
\usepackage{multirow}
\usepackage{fancyhdr}
\usepackage{eso-pic}

\AddToShipoutPictureBG*{%
  \AtPageUpperLeft{%
    \raisebox{-\baselineskip}{%
      \makebox[\paperwidth][c]{\textcolor{red}{Pre-print accepted in “9th International Conference on Engineering and Emerging Technologies (ICEET 2023)” IEEE Turkey Section}}
    }
  }
}

\def\BibTeX{{\rm B\kern-.05em{\sc i\kern-.025em b}\kern-.08em
    T\kern-.1667em\lower.7ex\hbox{E}\kern-.125emX}}
\newcommand{\myauthorrefmark}[1]{\textsuperscript{#1}}

\begin{document}

\title{SRSS: A New Chaos-Based Single-Round Single S-Box Image Encryption Scheme for Highly Auto-Correlated Data\\
}

\author{
\IEEEauthorblockN{
    Muhammad Shahbaz Khan\myauthorrefmark{1},
    Jawad Ahmad\myauthorrefmark{1},
    Hisham Ali\myauthorrefmark{1},
    Nikolaos Pitropakis\myauthorrefmark{1}
}
\IEEEauthorblockN{
    Ahmed Al-Dubai\myauthorrefmark{1},
    Baraq Ghaleb\myauthorrefmark{1},
    William J. Buchanan\myauthorrefmark{1}
}
\IEEEauthorblockA{\myauthorrefmark{1}\textit{School of Computing, Engineering and the Built Environment}, \\
\textit{Edinburgh Napier University}, \\
Edinburgh, UK, \\
Emails: \{muhammadshahbaz.khan, j.ahmad, h.ali, n.pitropakis, a.al-dubai, B.Ghaleb, b.buchanan\}@ napier.ac.uk}
}

\maketitle

\begin{abstract}
With the advent of digital communication, securing digital images during transmission and storage has become a critical concern. The traditional s-box substitution methods often fail to effectively conceal the information within highly auto-correlated regions of an image. This paper addresses the security issues presented by three prevalent S-box substitution methods, i.e., single S-box, multiple S-boxes, and multiple rounds with multiple S-boxes, especially when handling images with highly auto-correlated pixels. To resolve the addressed security issues, this paper proposes a new scheme SRSS—the Single Round Single S-Box encryption scheme. SRSS uses a single S-box for substitution in just one round to break the pixel correlations and encrypt the plaintext image effectively. Additionally, this paper introduces a new Chaos-based Random Operation Selection System—CROSS, which nullifies the requirement for multiple S-boxes, thus reducing the encryption scheme's complexity. By randomly selecting the operation to be performed on each pixel, driven by a chaotic sequence, the proposed scheme effectively scrambles even high auto-correlation areas. When compared to the substitution methods mentioned above, the proposed encryption scheme exhibited exceptionally well in just a single round with a single S-box. The close-to-ideal statistical security analysis results, i.e., an entropy of 7.89 and a correlation coefficient of 0.007, validate the effectiveness of the proposed scheme. This research offers an innovative path forward for securing images in applications requiring low computational complexity and fast encryption and decryption speeds.
\end{abstract}
\vspace{10pt}
\begin{IEEEkeywords}
S-Box, chaos, image encryption, correlation, single round, single S-Box
\end{IEEEkeywords}

\section{Introduction}
\begin{figure}[b]
    \centering
    \includegraphics[width=0.9\linewidth]{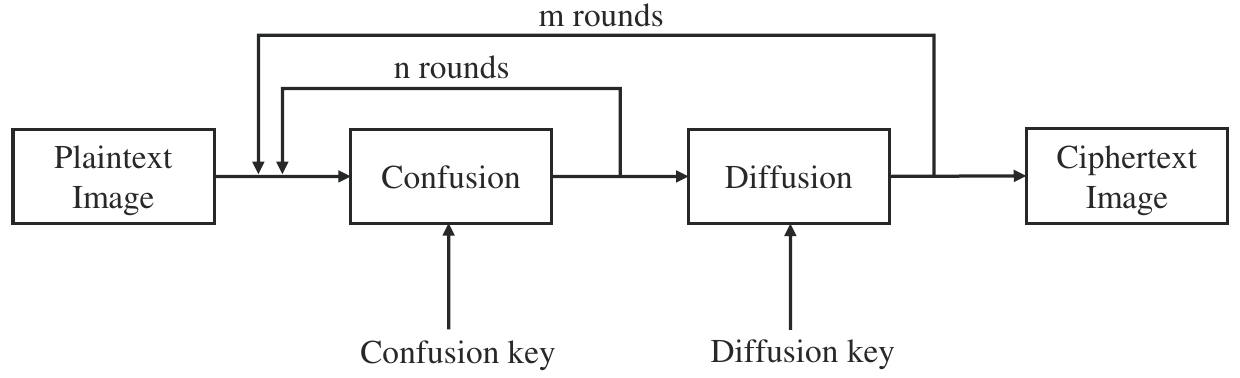}
    \caption{Image encryption basic workflow with confusion and diffusion processes.}
    \label{fig:fig1-image_encryption}
\end{figure}

With the rapid development of digital communication, social media, telemedicine (to transmit or store clinical image), online biometric systems (to store and transmit face portraits or fingerprints), and the Internet of Things, a large amount of digital images is transmitted over the internet and stored in cloud storage \cite{ahmed_2022_a}. The information in these digital images may be illegally intercepted, destroyed, or tampered with during transmission or storage \cite{sahu_2023_a, younes_2008_image}. Therefore, digital images need a high level of security. Image encryption plays an indispensable role in securing digital images. Image encryption involves two basic processes, i.e., confusion and diffusion. According to Claude Shannon \cite{shannon_1949_communication}, confusion refers to changing the values of the pixels based on a key and is usually achieved by substituting one value for another. Diffusion, on the other hand, refers to changing the position of the pixels based on a key. This is usually achieved through mechanisms like the permutation. The basic workflow of image encryption using confusion and diffusion processes is given in Fig. \ref{fig:fig1-image_encryption} and is mathematically expressed as follows [5]:
\begin{equation}
    C = \delta^n \left( \gamma^m (P, K_\delta), K_\gamma \right)
\end{equation}
where \( P \) is the plaintext image, \( C \) is the ciphertext image, \( \delta \) and \( \gamma \) represent the confusion and diffusion processes, respectively, \( K_\delta \) and \( K_\gamma \) are the confusion and diffusion secret keys, and \( n \) and \( m \) are the number of rounds for confusion and diffusion. A secure image encryption algorithm should be sensitive to the cipher key, with a larger key space to be effective against brute force and other attacks. The key space for a general image encryption system can be computed by Equation 2 \cite{ahmad_2015_a}.
\begin{equation}
KS = (KS_{\delta}^n \cdot KS_{\gamma})^m
\end{equation}
where \( KS_\delta \) and \( KS_\gamma \) represent the key spaces of the confusion and diffusion processes, respectively.

\begin{figure}[t]
    \centering
    \includegraphics[width=\linewidth]{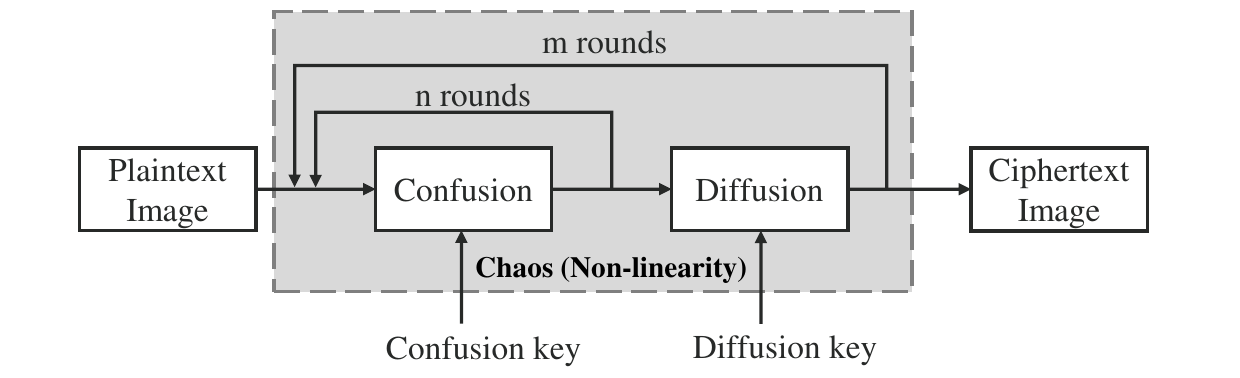}
    \caption{Image encryption with chaos-based confusion and diffusion processes.}
    \label{fig:fig2-image_encryption_chaos}
\end{figure}

Recently, chaos theory has proven to be an effective and efficient tool in image encryption, owing to its high sensitivity to initial conditions \cite{zhu_2019_a}, randomness \cite{wang_2023_an}, unpredictability \cite{ming_2022_a}, and ergodicity \cite{li_2023_the}. When combined with the confusion and diffusion processes in image encryption, it induces non-linearity in the encrypted image and significantly enhances the security of the encryption algorithm. Fig. \ref{fig:fig2-image_encryption_chaos} depicts where the chaos comes in an image encryption algorithm. In most cryptographic systems, the fundamental non-linear component of the confusion process is the S-Box (substitution box) \cite{yi_2019_a, liu_2017_chaos, chen_2018_exploiting}. The S-box substitution method transforms inputs into altered outputs.

Usually, three common types of S-box substitution methods are utilized: single S-box using bijective mapping \cite{anees_2014_chaotic, ahmad_2015_chaosbased, shafique_2020_dynamic}, multiple S-boxes \cite{khan_2015_a, wang_2013_a, hussain_2019_image, zhang_2018_efficient, zhu_2019_entropy, wang_2019_sbox}, and multiple rounds of encryption with multiple S-boxes \cite{zhou_2014_a, zhu_2012_a, ullah_2017_a}. However, a common drawback of these methods is their inability to handle images with high auto-correlation, where sections of similar pixel values simply transform into different brightness levels rather than becoming adequately encrypted. This issue has also been addressed and analyzed in detail in Section 2. To address these concerns, this paper aims at proposing a new image encryption scheme that effectively scrambles the image and also mitigates the computational and latency problems in existing schemes. The proposed scheme utilizes a single S-box and only a single round of substitution and breaks the correlations in the image, even in areas of high auto-correlation.

The main contributions of this paper are:
\begin{enumerate}
    \item A new image encryption scheme ‘SRSS – Single Round Single S-Box’ is proposed to resolve the security, complexity, and latency issues identified in traditional s-box substitution methods. This scheme breaks the correlations in the pixels and encrypts the image by utilizing a single S-box for substitution in only a single round.
    \item A new chaos-based random operation selection system – CROSS – is introduced, which eliminates the need for multiple s-boxes and hence, reduces the complexity of the encryption scheme.
    \item Three types of substitution methods, i.e., single s-box, multiple s-boxes, and multiple s-boxes with multiple rounds, have been implemented and analyzed to highlight the security issues, especially for images with highly auto-correlated pixels and lower gray scales.
\end{enumerate}

\section{Problem Formulation}
Three types of s-box substitution methods have been implemented and analyzed in detail to highlight their security issues.
\begin{figure}[t]
    \centering
    \begin{subfigure}[b]{0.6\linewidth}
        \includegraphics[width=\linewidth]{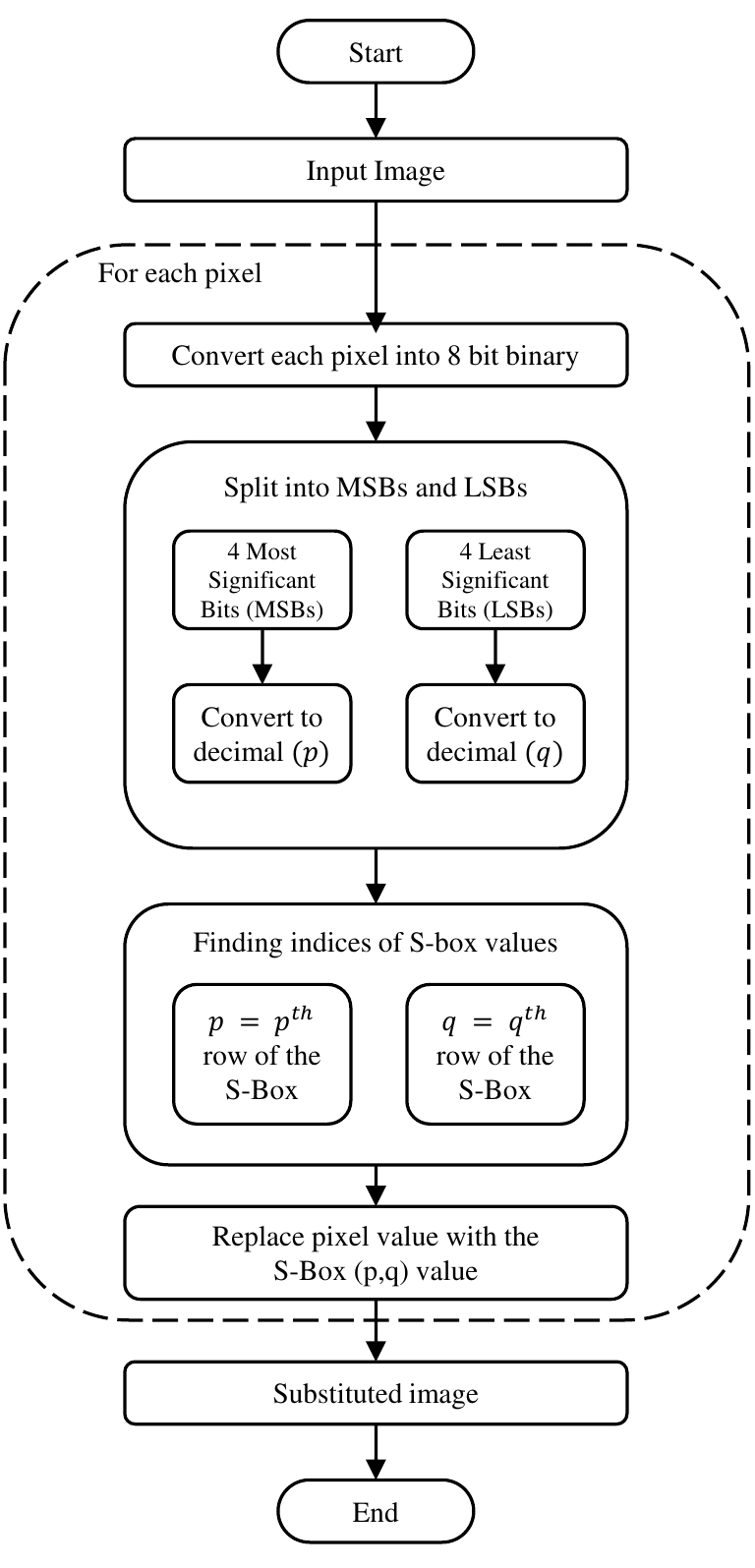}
        \caption{}
        \label{fig:fig3(a)bijective}
    \end{subfigure}
    
    \begin{subfigure}[b]{0.7\linewidth}
        \includegraphics[width=\linewidth]{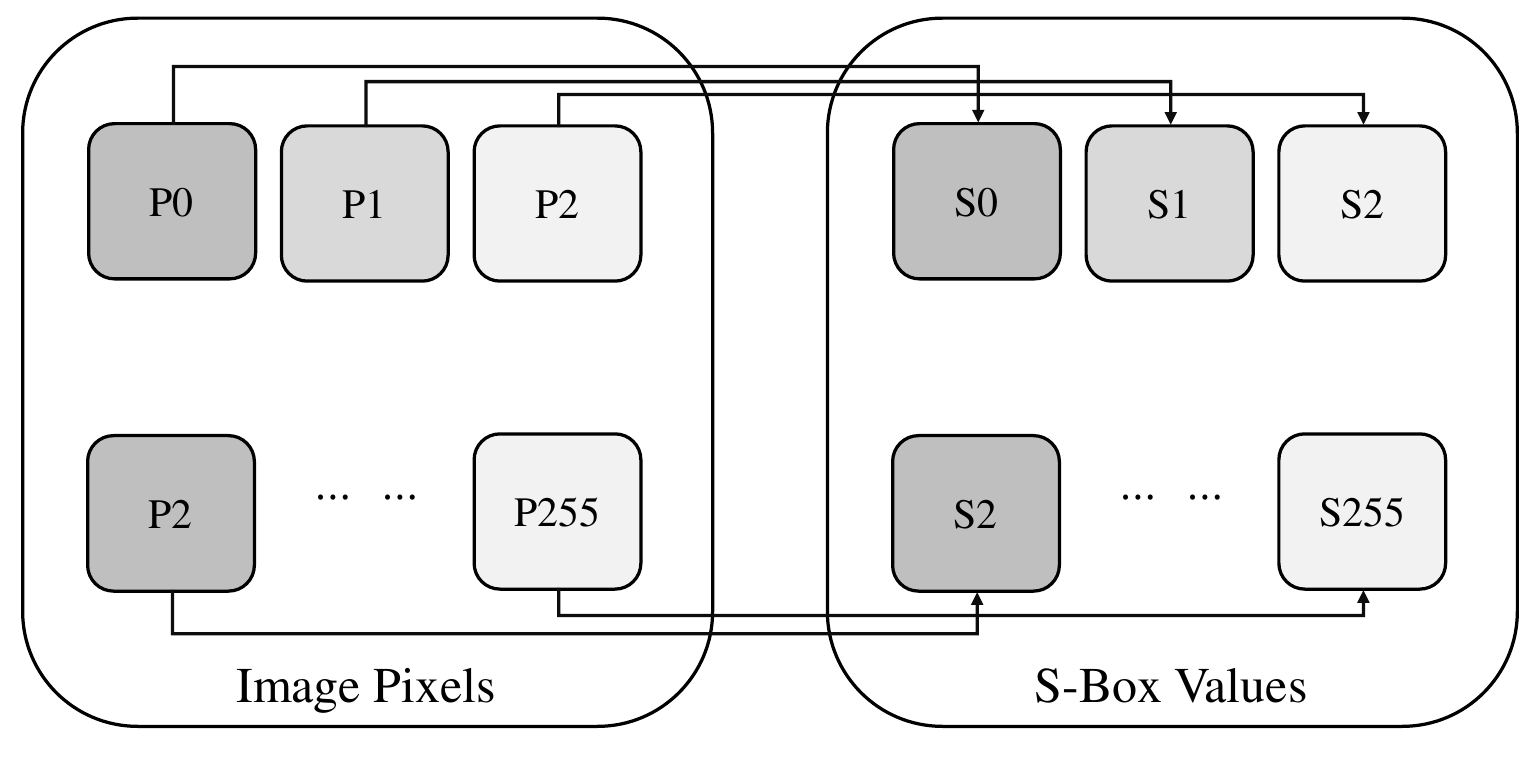}
        \caption{}
        \label{fig:fig3(b)single_S-box}
    \end{subfigure}
    \caption{(a) Single S-box substitution algorithm, (b) bijective mapping}
    \label{fig:fig3-single_sbox}
\end{figure}

\subsection{Single S-Box Substitution Method}
The substitution mapping used in single s-box substitution methods is called bijective mapping. In bijective mapping, pixels are replaced with only one unique S-box value, and the S-box is considered as the bijective function \(f(x)\). The substitution algorithm that utilizes bijective mapping is given in Fig. 3 (a) and the S-box bijective substitution function is given in Fig. 3 (b). This function can be realized mathematically as:
\begin{equation}
S: \text{GF}(2^p) \rightarrow \text{GF}(2^q)
\end{equation}
if \(x_1 = x_2\), then
\begin{equation}
    f(x_1)=f(x_2)
\end{equation}

In such s-box substitution function, the image is encrypted with only one unique element of the utilized S-Box. Pixels having identical values will be replaced with the same unique number from the S-Box and hence, will result in a change in the brightness level of the region only. The results of the single s-box substitution algorithm given in Fig. 4 show that the Coins image is not scrambled efficiently and all edges are visible.

\subsection{Multiple S-Box Substitution Method}
The most commonly used multiple S-Box substitution method is shown in Fig. 5. Here, chaos is used in conjunction with multiple S-boxes. Chaotic sequences are generated by using logistic map, which is given in Equation (5).
\begin{figure}
    \centering
    \begin{subfigure}{0.42\linewidth}
        \includegraphics[width=\linewidth]{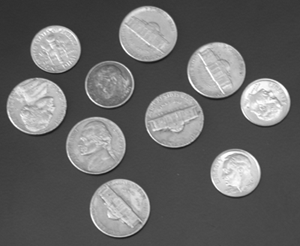}
        \caption{}
        \label{fig:sub1}
    \end{subfigure}
    \hfill
    \begin{subfigure}{0.45\linewidth}
        \includegraphics[width=\linewidth]{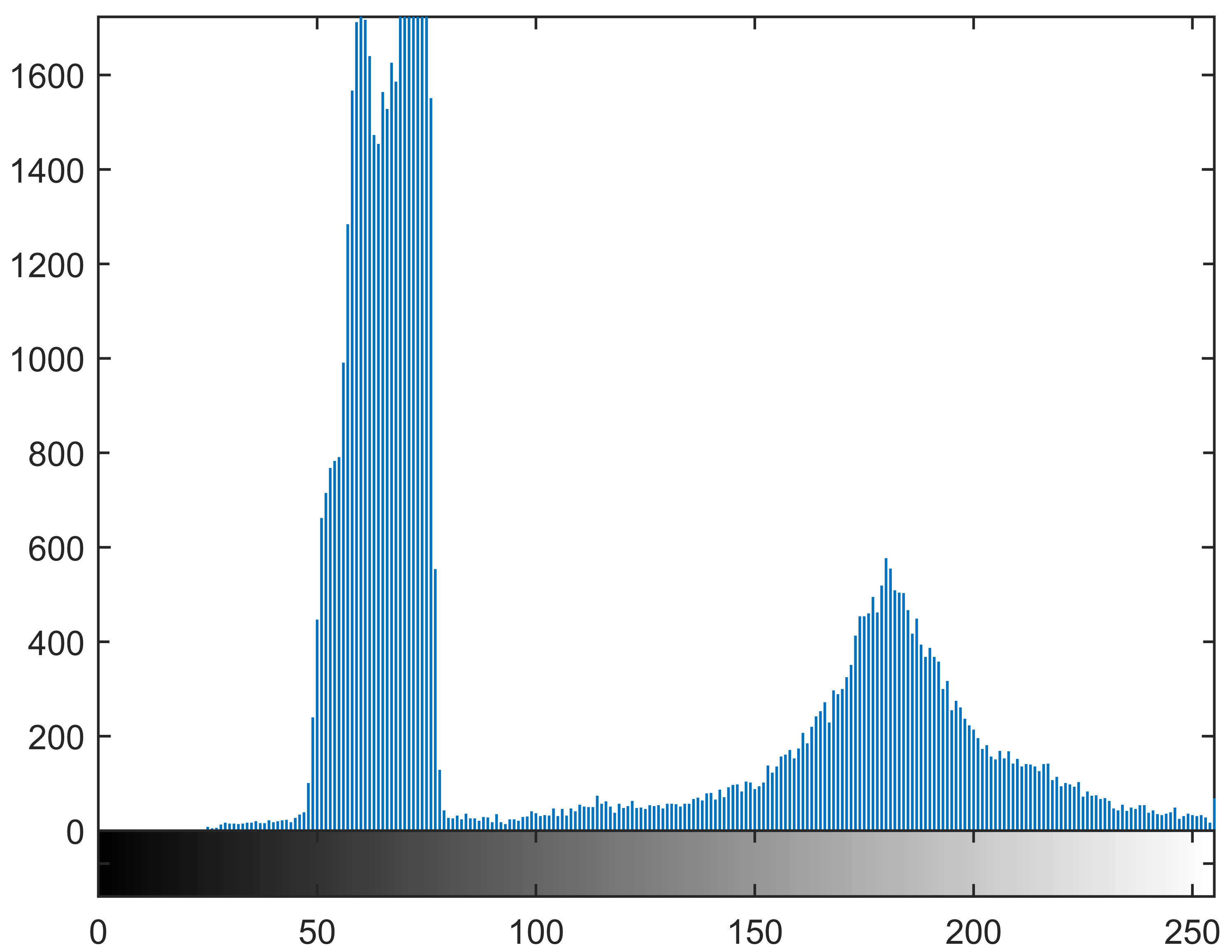}
        \caption{}
        \label{fig:sub2}
    \end{subfigure}
    \begin{subfigure}{0.42\linewidth}
        \includegraphics[width=\linewidth]{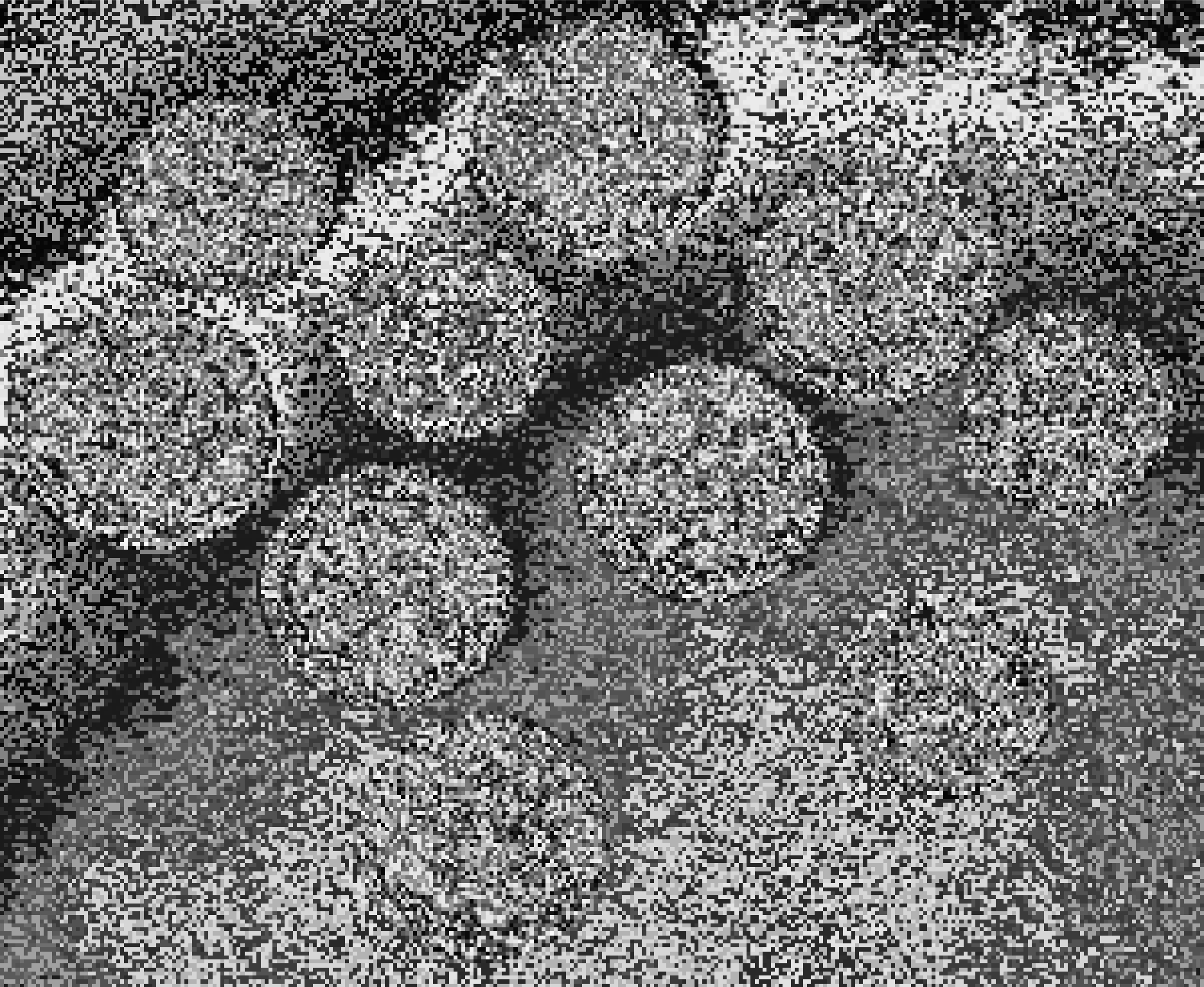}
        \caption{}
        \label{fig:sub3}
    \end{subfigure}
    \hfill
    \begin{subfigure}{0.45\linewidth}
        \includegraphics[width=\linewidth]{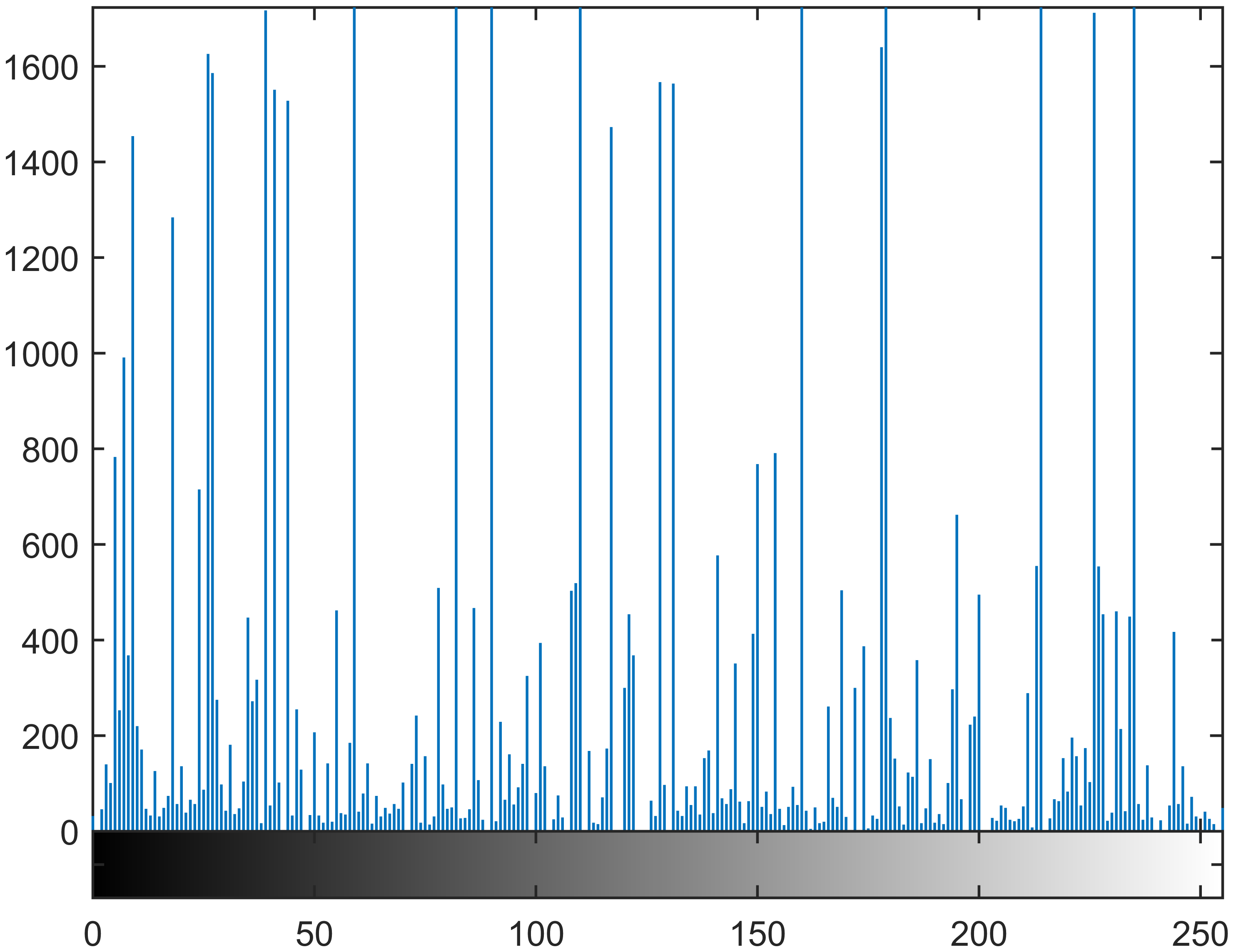}
        \caption{}
        \label{fig:sub4}
    \end{subfigure}
    \caption{Single s-box substitution results; (a-b) Coins image with its histogram, (c-d) Encrypted Coins image with its histogram.}
    \label{fig:fig4-single_s_box_results}
\end{figure}
\begin{equation}
    x_{n+1} = \mu \cdot x_n \cdot (1 - x_n)
\end{equation}
where \( \mu \in (0,4) \) and \( x_0 \in (0,1) \).

This scheme somehow resolves the issues of single s-box substitution, but the problem of visible edges continues to exist and is evident from the results shown in Fig. 6.

\subsection{Multiple S-Boxes and Multiple Rounds of Encryption}
\begin{figure}[t]
    \centering
    \includegraphics[width=0.9\linewidth]{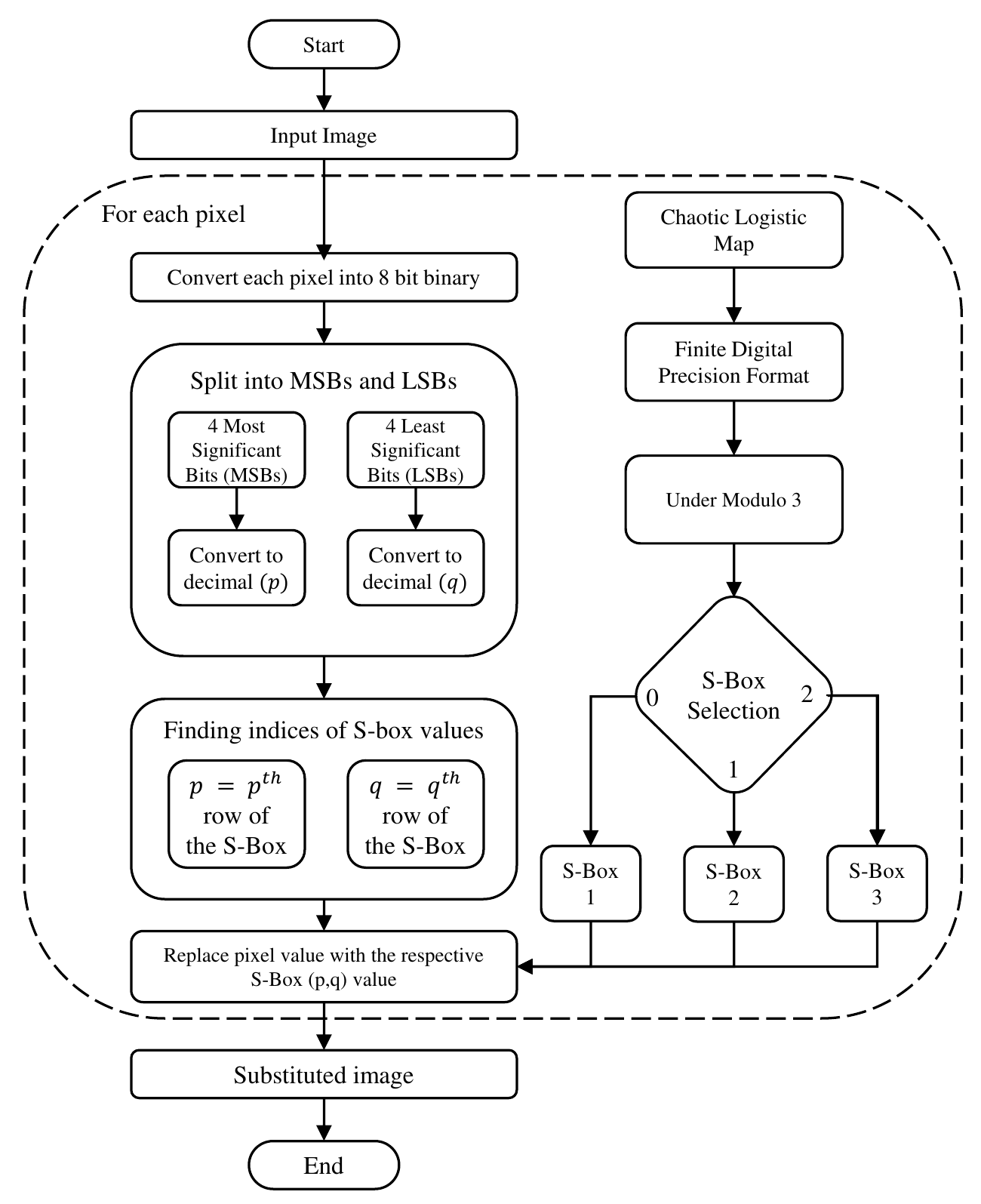}
    \caption{Multiple S-Box Chaotic Substitution Algorithm.}
    \label{fig:fig5-multiple_s_box}
\end{figure}
\begin{figure}[!ht]
    \centering
    \begin{subfigure}{0.42\linewidth}
        \includegraphics[width=\linewidth]{coins.png}
        \caption{}
        \label{fig:sub1}
    \end{subfigure}
    \hfill
    \begin{subfigure}{0.45\linewidth}
        \includegraphics[width=\linewidth]{single1b.png}
        \caption{}
        \label{fig:sub2}
    \end{subfigure}
    \begin{subfigure}{0.42\linewidth}
        \includegraphics[width=\linewidth]{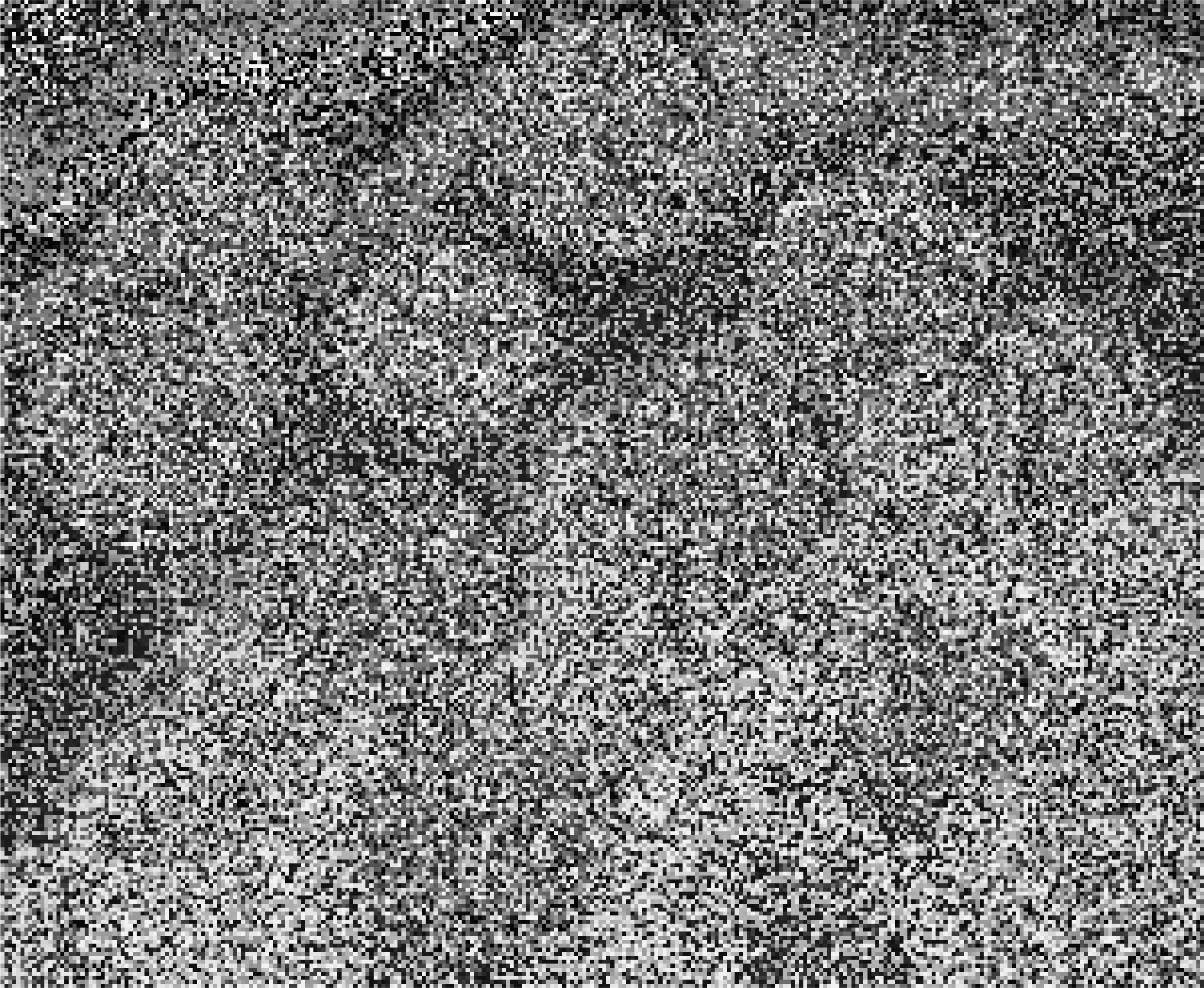}
        \caption{}
        \label{fig:sub3}
    \end{subfigure}
    \hfill
    \begin{subfigure}{0.45\linewidth}
        \includegraphics[width=\linewidth]{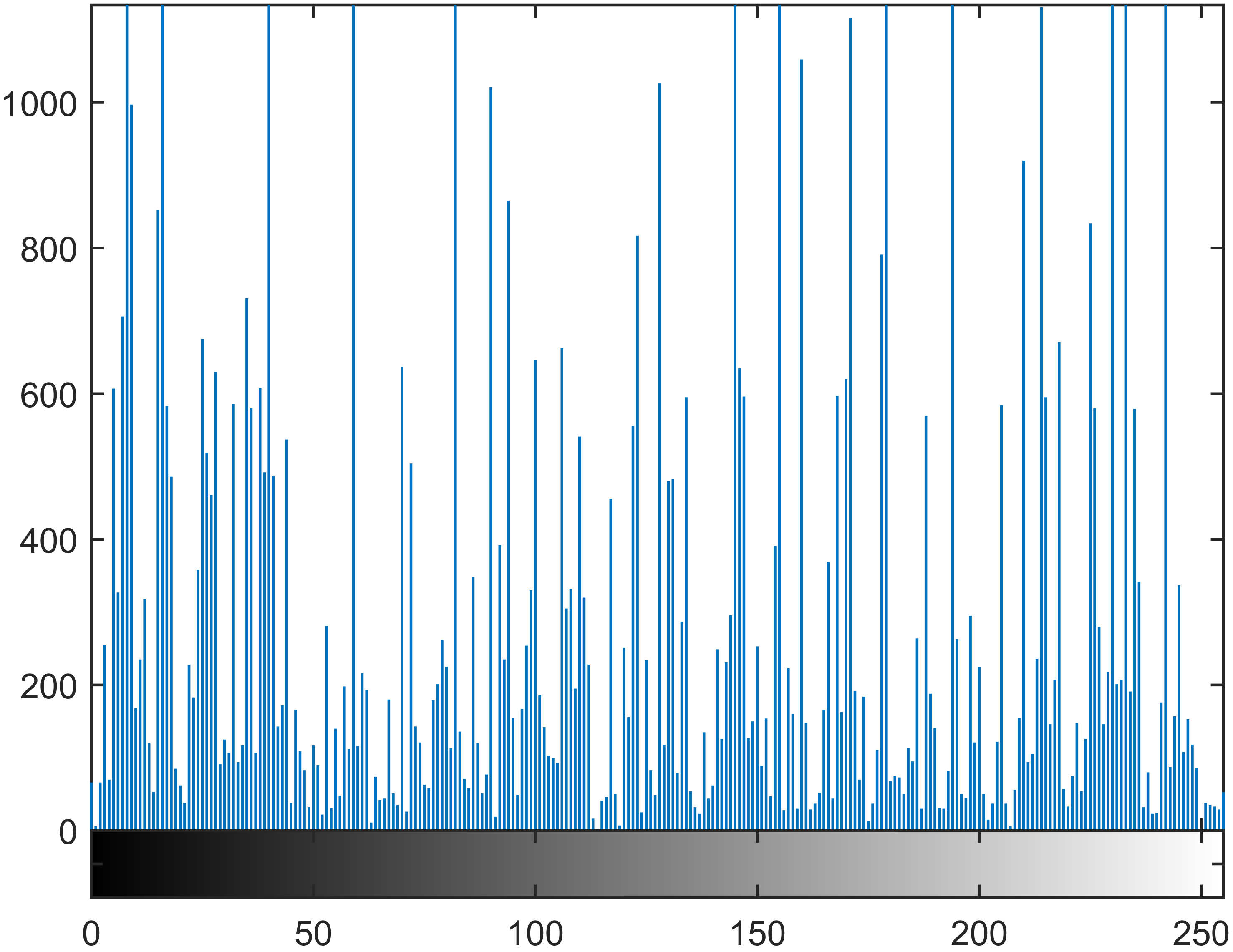}
        \caption{}
        \label{fig:sub4}
    \end{subfigure}
    \caption{Multiple s-box substitution results; (a-b) Coins image with its histogram, (c-d) Encrypted Coins image with its histogram.}
    \label{fig:fig4-single_s_box_results}
\end{figure}

In addition to single s-box and multiple s-box substitution methods, multiple rounds with multiple s-boxes-based methods are also utilized. We analyzed this method for 5 rounds of substitution and used three different s-boxes for substitution. It can be seen from the results in Fig. 7 that this method also fails to scramble the pixels effectively. Furthermore, the statistical security analysis in Table 1 also shows that there's no change in the entropy of the encrypted images after every substitution round. The results of GLCM (Gray Level Co-occurrence Matrix) parameters, i.e., correlation, contrast, energy, and homogeneity are almost the same after all rounds.

\subsection{Problem Statement}
In traditional S-box substitution methods, information within highly auto-correlated regions is not adequately concealed, i.e., the areas where pixel values are identical, such as sharp edges in an image. The fact that edges remain highly visible raises significant security concerns about the effectiveness of such substitution methods. 

\begin{figure}[t]
    \centering

    \begin{subfigure}[b]{\linewidth}
        \centering
        \includegraphics[width=0.9\linewidth]{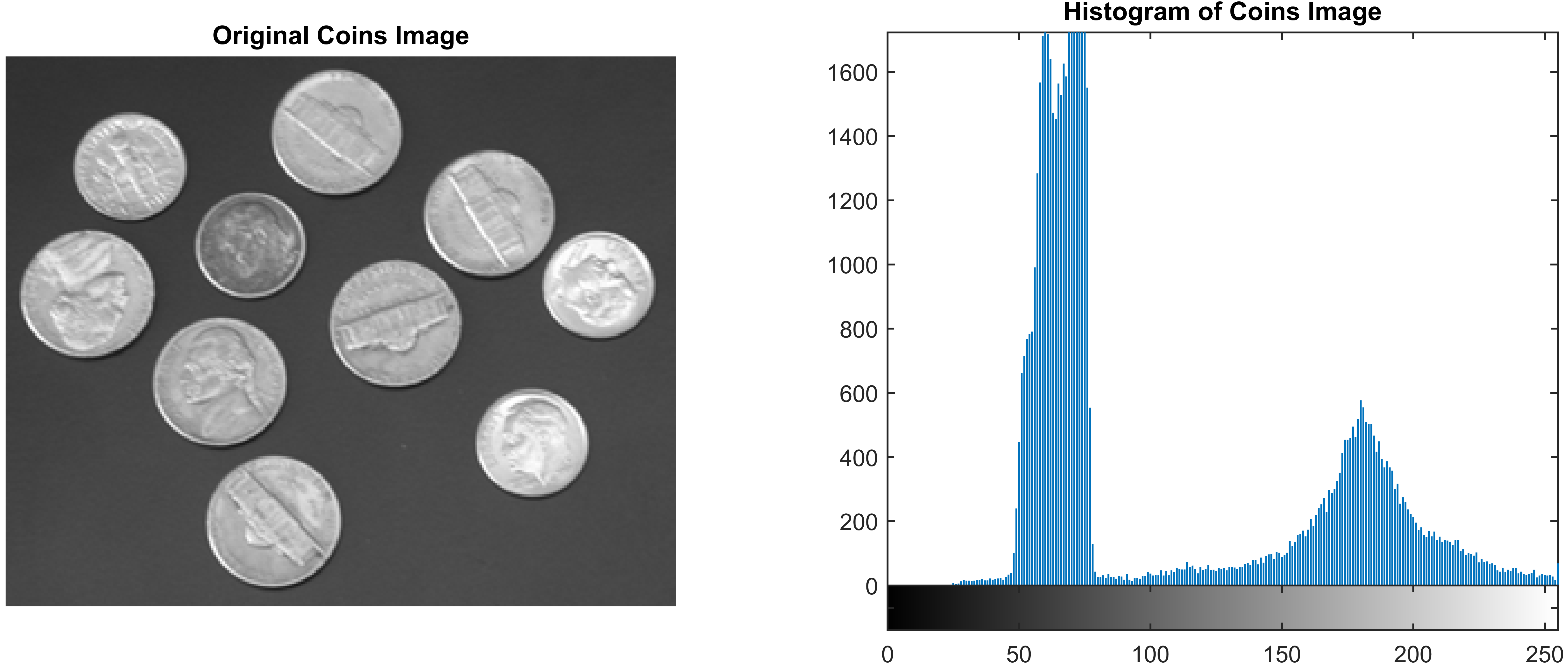}
        \caption{}
    \end{subfigure}

    \begin{subfigure}[b]{\linewidth}
        \centering
        \includegraphics[width=0.9\linewidth]{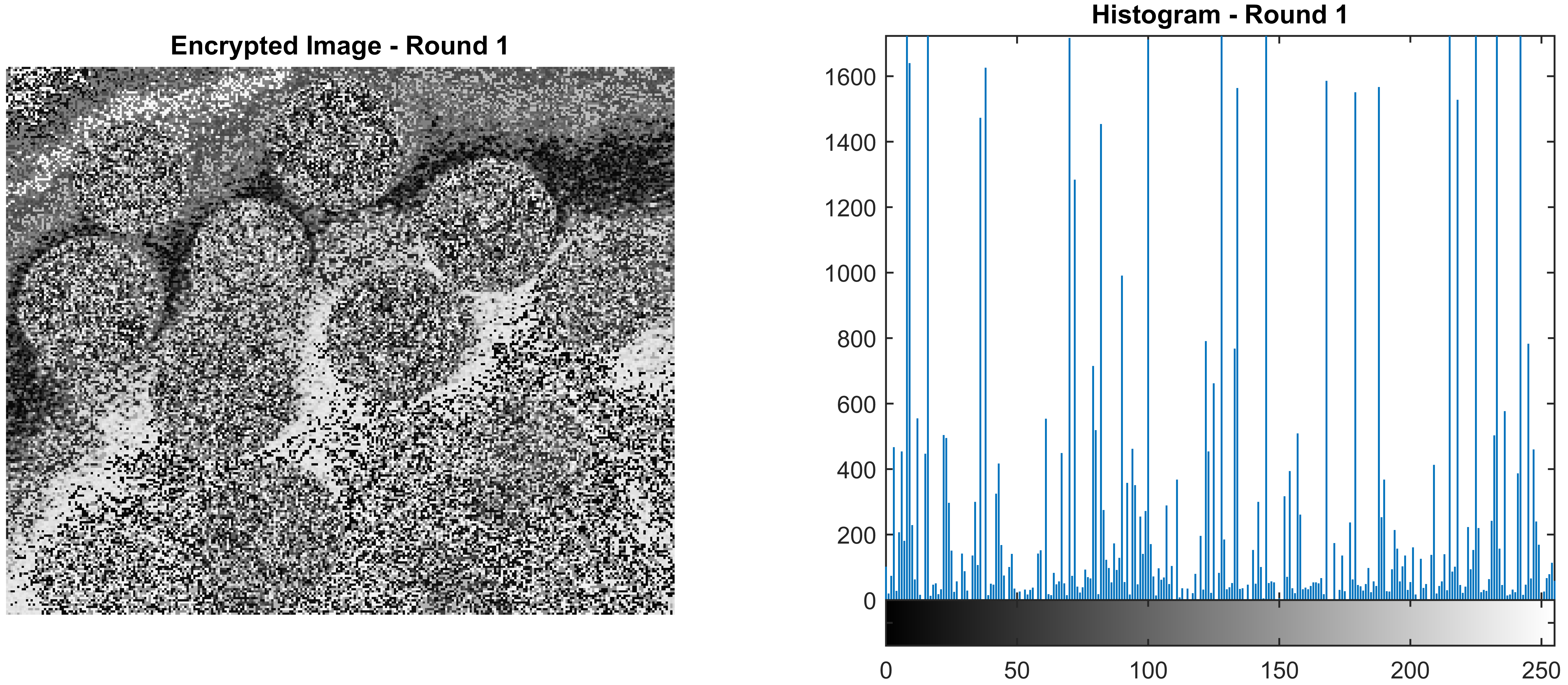}
        \caption{}
    \end{subfigure}

    \begin{subfigure}[b]{\linewidth}
    \centering
    \parbox[c]{\linewidth}{\centering\vdots}
    \end{subfigure}

    \begin{subfigure}[b]{\linewidth}
        \centering
        \includegraphics[width=0.9\linewidth]{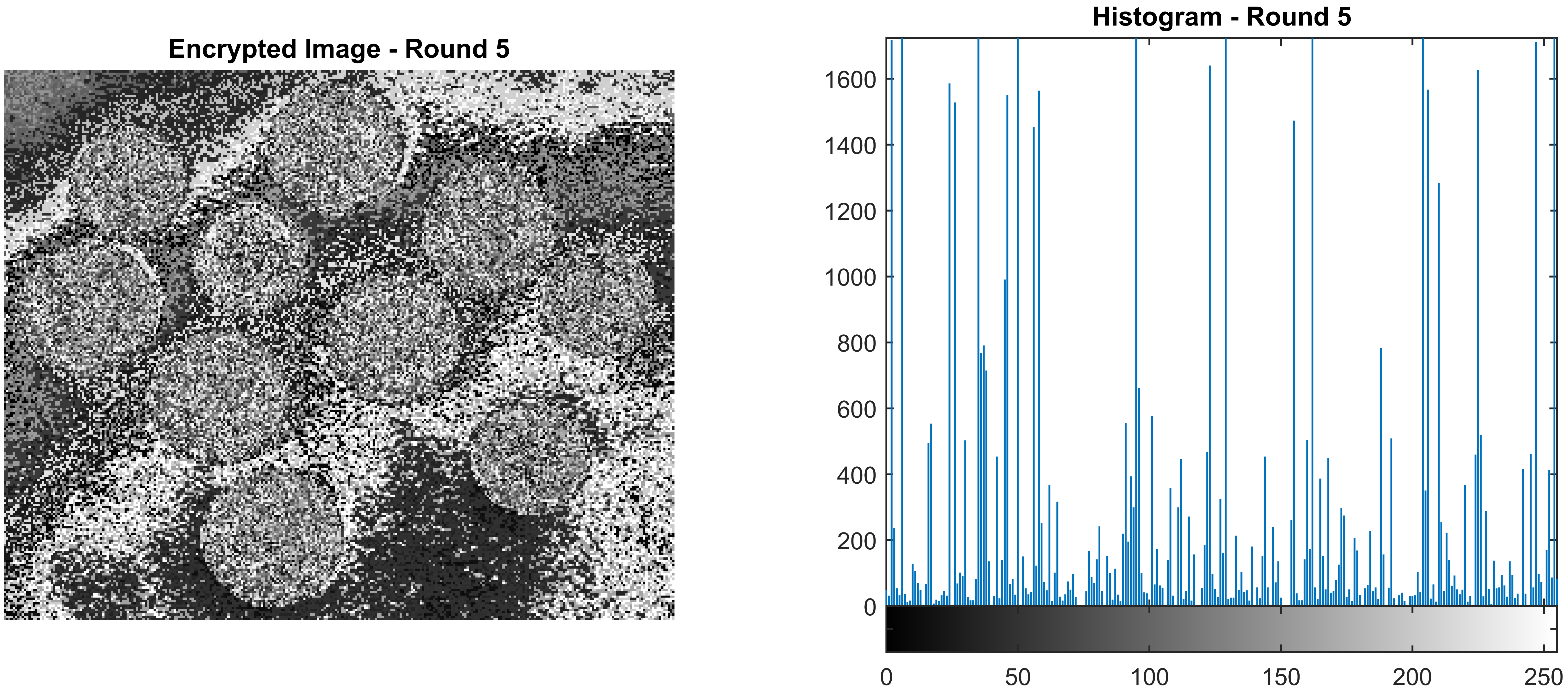}
        \caption{}
    \end{subfigure}

    \caption{Results of multiple rounds substitution showing no significant improvement; (a) Plaintext image, (b-c) Rounds 1 to 5.}
    \label{fig:multiple_rounds_results}
\end{figure}

This paper focuses on creating a substitution method based on a single round and single S-box that effectively scrambles the pixels of a plaintext image, eliminating the need for multiple rounds and S-boxes. Such methods are advantageous in applications demanding low computational complexity and faster encryption and decryption speeds.

\section{The Proposed Encryption Scheme}
The proposed encryption scheme utilizes a single S-box and only a single round of substitution. Each pixel value is replaced by a value from the S-box, but before substitution, it undergoes a randomly selected operation. The proposed scheme is explained in two parts: (a) SRSS - Single Round Single S-box Encryption Scheme, and (b) CROSS – Chaos-based Random Operation Selection System. The SRSS represents the entire encryption scheme. The CROSS, on the other hand, entails the random operation selection component of the scheme.
\begin{figure}[t]
    \centering
    \includegraphics[width=0.8\linewidth]{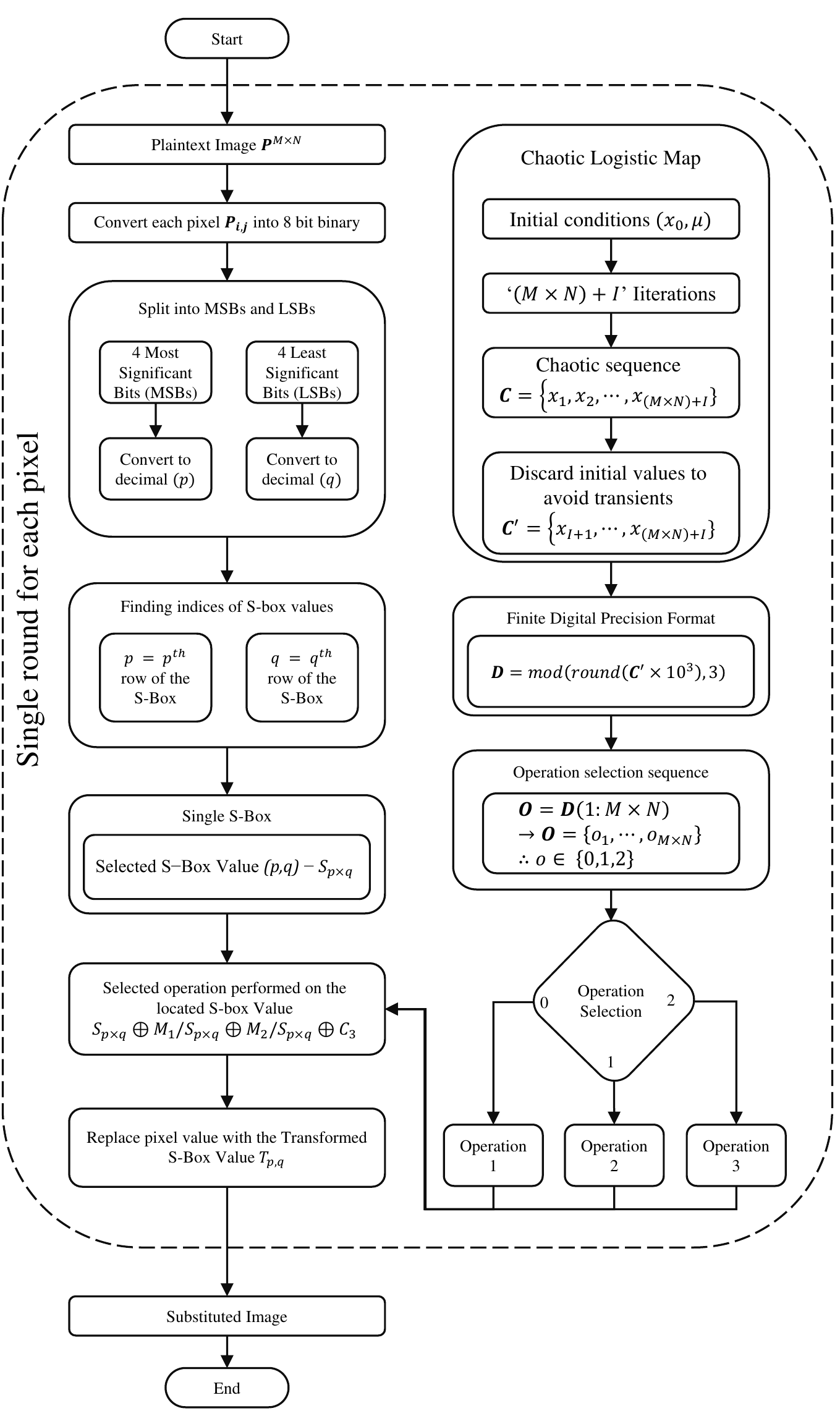}
    \caption{SRSS –The proposed single-round single S-box encryption scheme.}
    \label{fig:fig9-SRSS}
\end{figure}
\subsection{SRSS – Single-Round Single S-box Encryption Scheme}
The complete steps involved in the proposed SRSS encryption scheme, depicted in Fig. 8, are as follows.
\begin{itemize}
    \item \textbf{Step 1:} Input the plaintext image \( P^{(M \times N)} \) – \( M \times N \) denoting the dimension of the plaintext image. Also, initiate the secret keys for the chaotic map \((\mu,x_0)\), i.e., the control parameter and the initial condition, analyzed in Section 2.2.
    
    \item \textbf{Step 2:} Iterate the chaotic map equation '\( (M \times N) + I \)' times to generate a chaotic sequence \( C=\{x_1,x_2, \cdots, x_{(M \times N)+I} \} \). Here, \( I \) is the number of initial iterations to be discarded to avoid transients.
    
    \item \textbf{Step 3:} Discard the initial \( I \) iterations to avoid transients and keep the last \( M \times N \) values, i.e., \( C' = \{ x_{I+1}, \cdots, x_{(M \times N)+I} \} \).
    
    \item \textbf{Step 4:} The generated chaotic sequence \( C' \) has fractional values between 0 and 1. Apply finite digital format to convert these fractional values to a sequence of integers \( D \), i.e., \( D = \text{mod}(\text{round}(C' \times 10^3 ),3) \).

   \begin{table*}[!ht]
    \caption{Statistical Security Analysis of Multiple Rounds Substitution}
    \begin{center}
    \begin{tabular}{|p{2.5cm}|c|c|c|c|c|c|}
        \hline
        \multicolumn{2}{|c|}{\textbf{Security Parameter}} & \textbf{Round 1} & \textbf{Round 2} & \textbf{Round 3} & \textbf{Round 4} & \textbf{Round 5} \\
        \hline
        & Entropy & 6.316 & 6.316 & 6.316 & 6.316 & 6.316 \\
        \hline
        \multirow{4}{=}{GLCM} & Contrast & 10.57 & 7.89 & 10.63 & 8.41 & 9.32 \\
        \cline{2-7}
        & Correlation & 0.144 & 0.199 & 0.126 & 0.194 & 0.250 \\
        \cline{2-7}
        & Energy & 0.025 & 0.025 & 0.037 & 0.025 & 0.032 \\
        \cline{2-7}
        & Homogeneity & 0.48 & 0.51 & 0.51 & 0.51 & 0.52 \\
        \hline
    \end{tabular}
    \label{table:table1}
    \end{center}
\end{table*}
    
    \item \textbf{Step 5:} The modulus 3 operation in Step 3 makes sure that the chaotic sequence contains the values 0, 1, and 2. This makes our operation selection sequence of \( M \times N \) dimension, i.e., \( O = D(1:M \times N) \rightarrow O = \{ o_1, \cdots, o_{M \times N} \} \) such that \( o \in \{ 0,1,2 \} \).
    
    \item \textbf{Step 6:} Convert each pixel of the plaintext image \( P_{i,j} \) into 8-bit binary and split the 8-bit binary into two equal parts, making the first 4 bits the Most Significant Bits (MSBs) and the last 4 bits the Least Significant Bits (LSBs).
   
    \item \textbf{Step 7:} To find the indices of the S-box values, which will replace the pixel of the plaintext image, convert the MSBs to decimal \( p \) and LSBs to decimal \( q \). \( p \) corresponds to the row number of the S-box and \( q \) corresponds to the column number of the S-box, locating the S-box value \( S_{p,q} \).
    
    \item \textbf{Step 8:} The operation selection sequence \( O \) containing values 0, 1, and 2, selects one of the three operations to be performed on the selected S-box value \( S_{p,q} \). 0 selects Operation 1, 1 selects Operation 2, and 2 selects Operation 3. This selection is random based on the value in the operation selection sequence \( O \).
    
    \item \textbf{Step 9:} The selected operation is performed on the selected S-box value \( S_{p,q} \) and converts it into a new transformed value \( T_{p,q} \). This transformed value then replaces the original pixel \( P_{i,j} \) in the plaintext image.
\end{itemize}

\subsection{CROSS – Chaos-based Random Operation Selection System}
The Chaos-based Random Operation Selection System makes sure that every time for each pixel, a random operation is selected from the three operations. The operation selection sequence \( O \) is generated via a chaotic logistic map and contains random values of \( 0 \), \( 1 \), and \( 2 \). \( 0 \) corresponds to operation 1, \( 1 \) corresponds to operation 2, and \( 2 \) corresponds to operation 3. For the sake of simplicity, the operation chosen for all three operations is Bit X-OR. Three modifier constants or CROSS- secret keys, i.e., \( M_1 \), \( M_2 \) and \( M_3 \) are chosen. \( M_1 \), \( M_2 \), and \( M_3 \) $\in \{0, \ldots, 255\}$. In operation 1, the selected S-box value is first Bit XORed with \( M_1 \) before replacing the original pixel value of the plaintext image, similarly, in operations 2 and 3, the selected S-box value is bit XORed with \( M_2 \) and \( M_3 \), respectively. The designed chaos-based random operation selection system is given in Fig. 9.

\section{Results of the proposed SRSS scheme}
\subsection{Encryption Results of the Proposed Encryption Scheme}
The proposed SRSS exhibited effective confusion of the plaintext image in just one round. The random selection of operations performed on the selected S-box values ensured that no edges are visible and all pixels have been replaced with several distinct values. The SRSS encrypted image with its histogram is given in Fig. 10. Furthermore, the results of the statistical security analysis given in Table 2 showing close to ideal values of entropy and correlation also validated the effectiveness of the proposed encryption scheme.

\begin{figure}[t]
    \centering
    \includegraphics[width=0.7\linewidth]{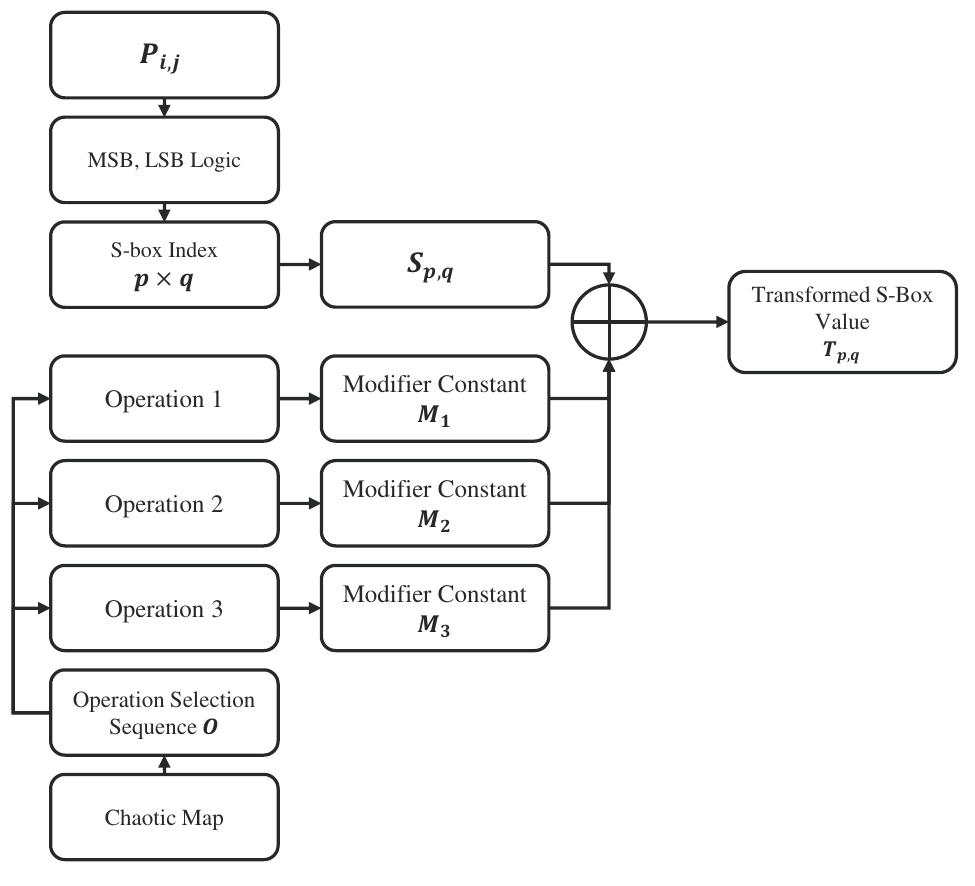}
    \caption{CROSS – The proposed chaos-based random operation selection system.}
    \label{fig:fig10-CROSS}
\end{figure}
\begin{table}[!bp]
    \caption{Statistical Security Analysis of the Proposed SRSS encryption scheme}
    \centering
    \begin{tabular}{|p{2.5cm}|p{2.5cm}|p{2.5cm}|}
        \hline
        \multicolumn{2}{|c|}{\textbf{Security Parameter}} & \textbf{Single Round} \\
        \hline
        & Entropy & 7.989 \\
        \hline
        \multirow{4}{=}{GLCM} & Contrast & 10.45 \\
        \cline{2-3}
        & Correlation & 0.0007 \\
        \cline{2-3}
        & Energy & 0.015 \\
        \cline{2-3}
        & Homogeneity & 0.389 \\
        \hline
    \end{tabular}
    \label{table:table1}
\end{table}

\subsection{Comparison with Multiple S-boxes and Multiple Rounds Algorithm}
When compared with the results of substitution methods under study, the proposed scheme exhibited considerably good security performance. It is evident from Fig. 11 that the proposed SRSS encryption scheme encrypts the plaintext image more effectively as compared to the round 5 encrypted image of the multiple s-box system.

\section{Conclusion}
To resolve the security, latency, and computational concerns associated with traditional S-box substitution methods, this paper addressed some inherent security vulnerabilities in three types of S-box substitution methods, especially when dealing with images that have highly auto-correlated pixels and lower gray scales. Furthermore, to resolve the highlighted security concerns, this paper proposed a robust Single Round Single S-Box (SRSS) encryption scheme that simplifies the encryption process while enhancing its security efficacy. In addition to the proposed SRSS, this paper introduced a new Chaos-based Random Operation Selection System (CROSS), a mechanism designed to reduce the complexity of the encryption scheme by negating the need for multiple S-boxes. The new methods demonstrated their potency by outperforming the existing substitution methods in terms of statistical security analysis. The SRSS and CROSS collectively achieved near-ideal results with an entropy of 7.89 and a correlation coefficient of 0.007, thus substantiating their effectiveness.
\begin{figure}[h]
    \centering
    \begin{subfigure}{0.42\linewidth}
        \includegraphics[width=\linewidth]{coins.png}
        \caption{}
        \label{fig:sub1}
    \end{subfigure}
    \hfill
    \begin{subfigure}{0.45\linewidth}
        \includegraphics[width=\linewidth]{single1b.png}
        \caption{}
        \label{fig:sub2}
    \end{subfigure}
    \begin{subfigure}{0.42\linewidth}
        \includegraphics[width=\linewidth]{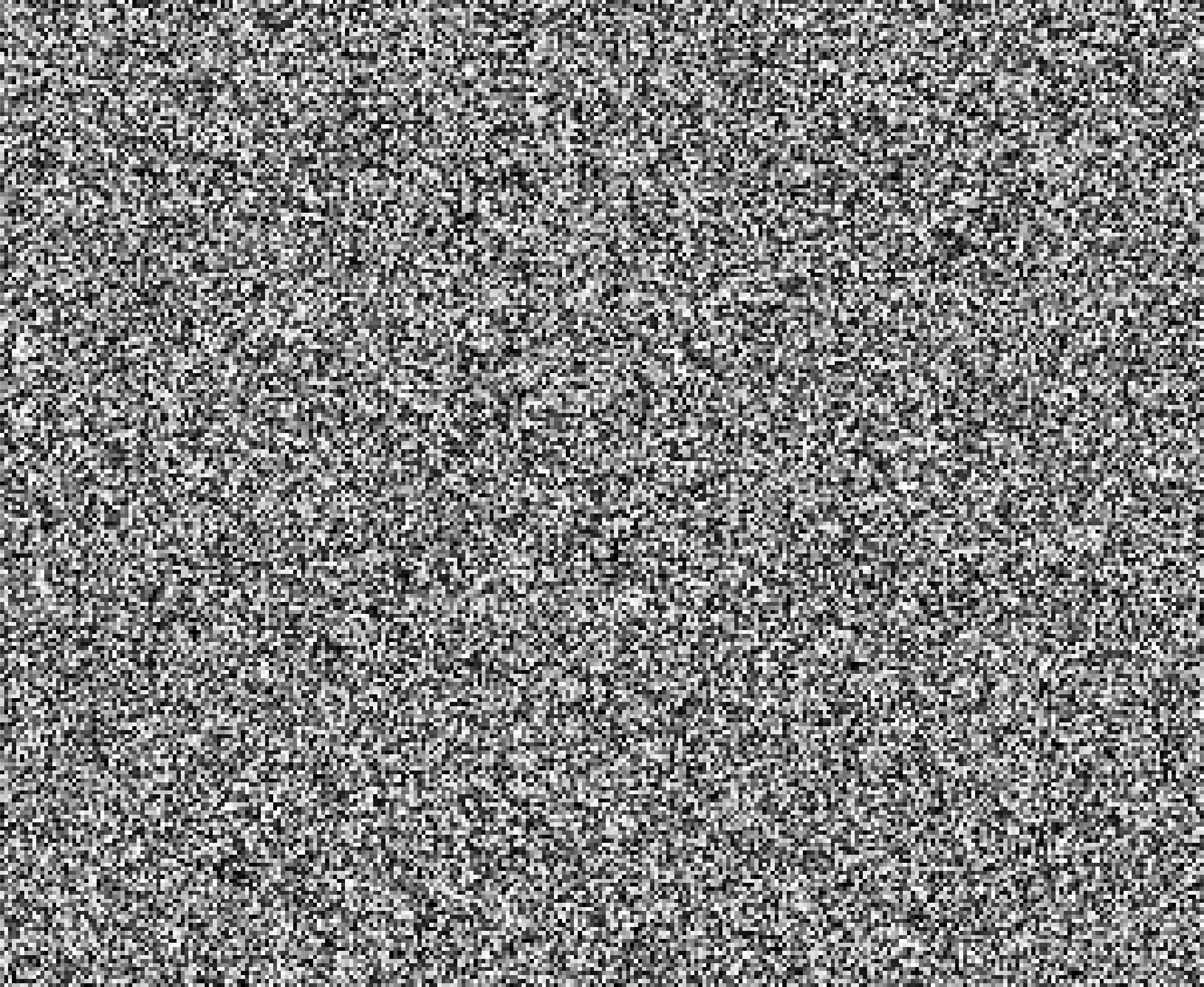}
        \caption{}
        \label{fig:sub3}
    \end{subfigure}
    \hfill
    \begin{subfigure}{0.44\linewidth}
        \includegraphics[width=\linewidth]{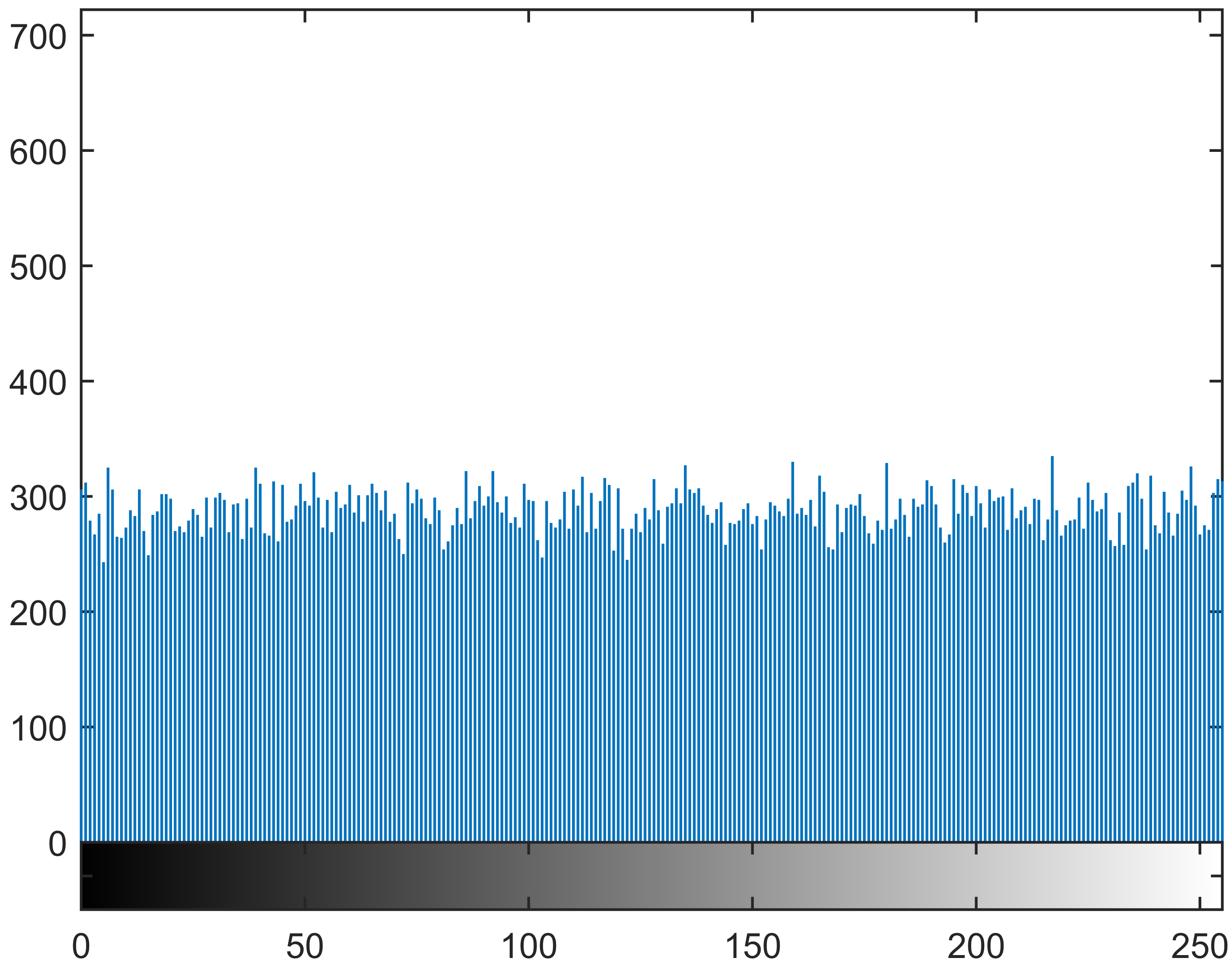}
        \caption{}
        \label{fig:sub4}
    \end{subfigure}
    \caption{Results of the proposed SRSS encryption scheme; (a-b) Coins image and its histogram, (c-d) SRSS Encrypted Coins image and its histogram.}
    \label{fig:fig4-single_s_box_results}
\end{figure}

\begin{figure}[h]
    \centering
    \begin{subfigure}{0.42\linewidth}
        \includegraphics[width=\linewidth]{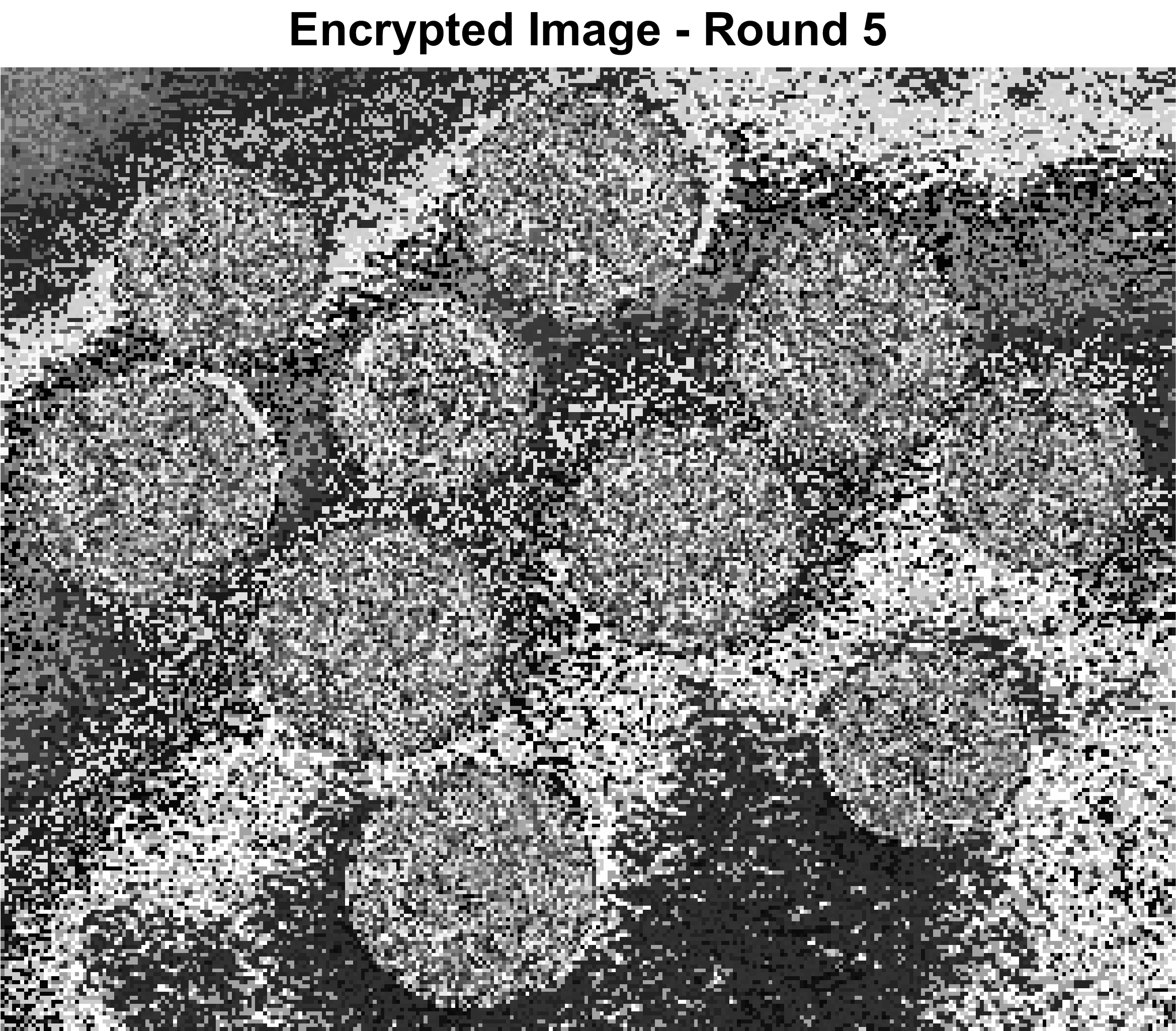}
        \caption{}
        \label{fig:sub1}
    \end{subfigure}
    \hfill
    \begin{subfigure}{0.42\linewidth}
        \includegraphics[width=\linewidth]{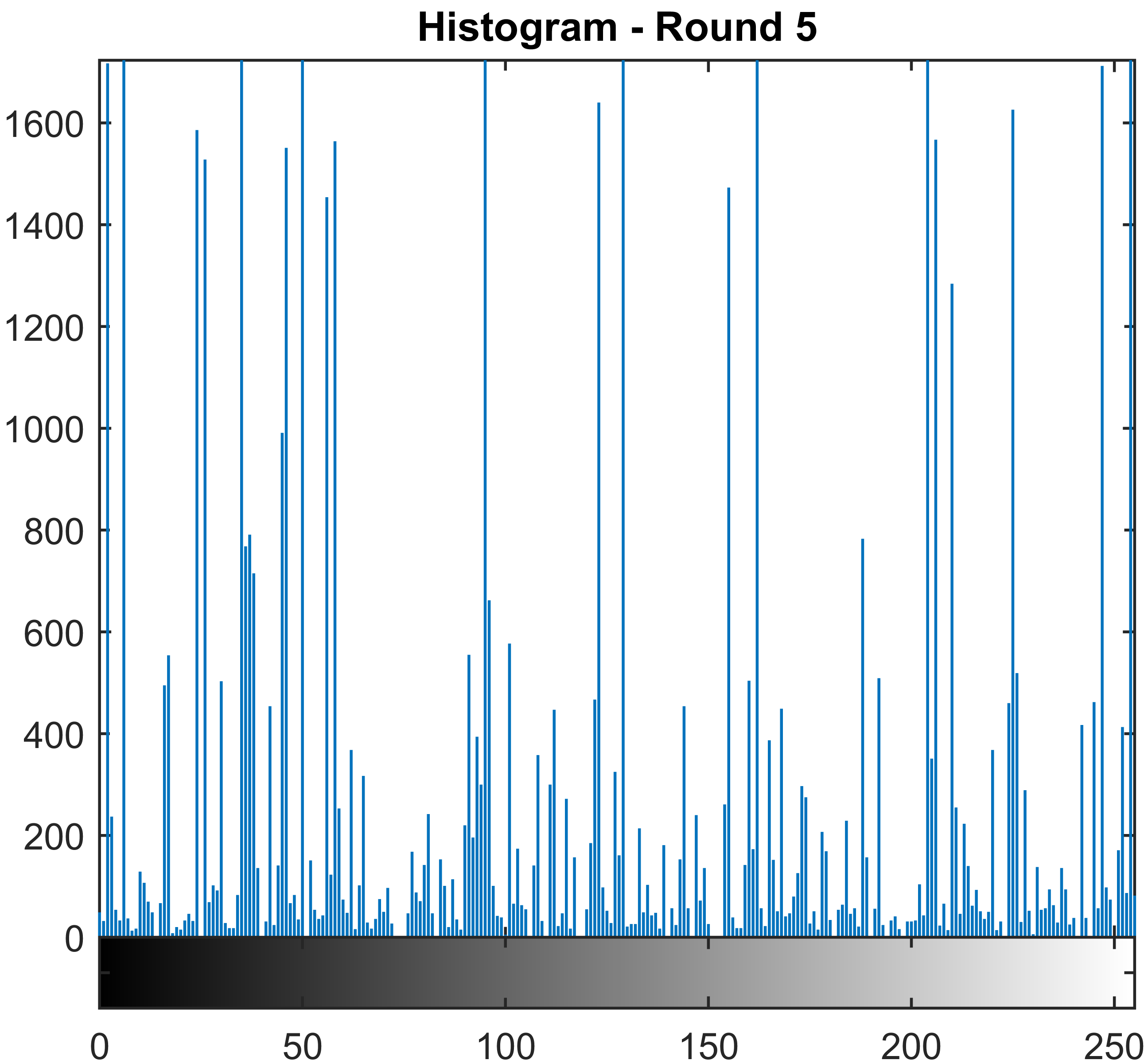}
        \caption{}
        \label{fig:sub2}
    \end{subfigure}
    \begin{subfigure}{0.42\linewidth}
        \includegraphics[width=\linewidth]{srss3.png}
        \caption{}
        \label{fig:sub3}
    \end{subfigure}
    \hfill
    \begin{subfigure}{0.44\linewidth}
        \includegraphics[width=\linewidth]{srss4.png}
        \caption{}
        \label{fig:sub4}
    \end{subfigure}
    \caption{. Comparison of the proposed SRSS encryption scheme; (a-b) Multiple rounds encrypted image and its Histogram, (c-d) The SRSS encrypted image and its Histogram.}
    \label{fig:fig4-single_s_box_results}
\end{figure}

\bibliographystyle{IEEEtran}
\bibliography{myreferences}

\end{document}